# CALCULATION OF REAL GAS PRESSURES USING MODIFIED FUNCTIONS OF PROBABILITY DISTRIBUTIONS

Liutuy A.I., Glushko L.N., Ostapenko A.A.*

Official address: Department of Physics, National Mining University, 19 K.Marks ave., Dnipropetrovsk, Ukraine.

* Corresponding author: postal address: Department of Physics, National Mining University, 19 K.Marks ave., Dnipropetrovsk, Ukraine
phone: +38 (093) 744-61-87
e-mail: ostapenko@optima.com.ua

We have improved the procedure for calculating distribution functions and the thermal equation of state for real gases by introducing modified functions of probability distributions into the gas phase state sum. Calculation of real gas pressure and the grand canonical partition function using the improved procedure demonstrate that numerical difficulties arising from a large number of state variables can be circumvented. We have also calculated real gas isotherms and probability of vaporization in phase equilibrium.

It has been long established that description of phase state and thermodynamics of critical phenomena are some of the fundamental problems in natural sciences.

The equation of state is one the most important properties of macroscopic isotropic matter. It is known that in the thermodynamic equilibrium volume $V$, pressure $p$ and temperature $T$ behave as $f(p,V,T)=0$ not only for ideal but also for real gases as well as for any other homogeneous and isotropic matter. It is not possible to derive the equation of state from general thermodynamic principles. In thermodynamics the equation of state is usually obtained from experimental data or taken from statistical mechanics. Past attempts to obtain the equation of state empirically have in essence been unsuccessful [1].

J.W. Gibbs laid the mathematical foundations for the phase transition theory introducing statistical principles into thermodynamics [1]. Fundamental laws of statistical physics are formulated in terms of Gibbs ensembles. In this communication we consider the phase transition "condensed phase-vapour". For this case the grand canonical ensemble is defined as the probability distribution of various states of quasi-isolated macroscopic system (real gas) that exchanges both energy and particles with the condensed phase (liquid). The calculation of a number of thermodynamic functions of real gas is based on the grand canonical partition function
$$Z = \sum_n \sum_j \exp\bigl((\mu n - \varepsilon_j)/kT\bigr)\cdot g(\varepsilon_j, n), \qquad (1)$$

where $\varepsilon_j$ is the energy of $j$-th energy level;
$g(\varepsilon_j, n)$ is the level's degeneracy or the statistical weight; $\mu$ is the chemical potential and $n$ is the number of particles in the gas phase.

When the energy spectrum of the molecule is known then the partition function can be used to calculate various thermodynamic functions of real gas. Pressure for example is expressed as
$$p = kT\cdot(\partial \ln Z/\partial V)_T. \qquad (2)$$

Although Gibbs outlined general mathematical principles for description of phase transitions in a macroscopic system, he only applied it to the simplest model of matter, that of ideal gas. His model of ideal gas presented particles void of any internal structure and where energy of atoms arises from their translation degrees of freedom. Using the definition of the partition function
$$Z = \frac{q_1^N}{N!} = \left[\exp\!\left(\frac{\psi_1}{kT}\right)(2\pi m kT)^{3/2} V\right]^N \frac{1}{N!}$$

(where $q_1$ is the probability coefficient; $m$ is the mass of the particle and $N$ is the number of particles) and equation (2) the pressure of ideal gas is:
$$p = N kT/V.$$

The theory has not been similarly applied to the case of real gas. Phase transitions in real gas occur with absorption of energy during evaporation of liquids and release of energy when vapour is condensed back into liquid. The process is characterised by parameter $\varepsilon_0$ – heat of vaporization.

It has been shown [2-4] that Gibbs statistical mechanics can be used to obtain the correct grand canonical partition function for real gases. The expression for probability of having $n$ molecules (out of the total number of molecules $N$) in the gas phase with energy $\varepsilon_j$

$$w_{\varepsilon_j,n} = \exp((\mu n - \varepsilon_j)/kT) \cdot \hat{O}(\varepsilon_j, n)/Z \quad (3)$$

has been modified to introduce the four fundamental parameters that describe the properties of the matter, namely critical pressure $p_c$, specific volume $v_c$, temperature $T_c$, and heat of vaporization $\varepsilon_0$. When we introduce specific thermodynamic parameters $\pi = p/p_c$, $\alpha = v/v_c$, $\tau = T/T_c$, and also express $\varepsilon_0$ using parameter $r = \varepsilon_0/(kT_c)$, and take the sum in (1) over the energy levels of molecule $\varepsilon_j$ we obtain:

$$Z = \sum_{n=0}^{N} \left[ ((N\alpha - n + 1)/(n+1)) \cdot \tau^{i/2} \cdot \exp(r(1 - \tau^{-1})) \right]^n, \quad (4)$$

where $i$ is the number of active degrees of freedom of the molecule.

When we factor out the term that depends on the specific temperature $\tau$, but not on the number of particles $n$:

$$A(\tau) = \tau^{i/2} \cdot \exp(r(1 - \tau^{-1})), \quad (5)$$

then the partition function becomes

$$Z = \sum_{n=0}^{N} q_n^n = \sum_{n=0}^{N} \left[ ((N\alpha - n + 1)/(n+1)) \cdot \dot{A}(\tau) \right]^n, \quad (6)$$

The authors of [5] have used expressions (2) and (6) to obtain the specific thermal equation of state of real gas:

$$\pi = K\tau \sum_{n=0}^{N} \left[ (n/(N\alpha - n + 1)) q_n^n \right] / Z, \quad (7)$$

where $K = RT_c/(p_c v_c M)$ is the critical coefficient [6]. Here $M$ is the molar mass of the matter and $R$ is the gas constant.

The probability of having $n$ molecules out of the total $N$ in the gas phase as expressed in (3) is

$$w(n) = q_n^n/Z =$$

$$= \sum_{n=0}^{N} \left[ \frac{N\alpha - n + 1}{(n+1)} \cdot \tau^{i/2} \cdot \exp(r(1 - \tau^{-1})) \right]^n / Z. \quad (8)$$

The above entity satisfies the property of mathematical probability $0 \leq w(n) \leq 1$. It can be seen that $0 \leq n \leq N$.

Fig.1 shows isotherms of real gas at and around the critical point on the logarithmic scale. The graph demonstrates that equation (7) is in full agreement with the experimental data and correctly describes the dependence of pressure on volume in the isothermic process both in the single phase (ideal gas), and in the two-phase (horizontal part of the isotherm) regimes, where a phase transition between condensed and gas phases of the matter occurs. Before the critical point ($\tau < \tau_c$, $A(\tau) < 1$) $\pi < 1$, and after

$$(\tau > \tau_c, A(\tau) > 1) \, \pi > 1.$$

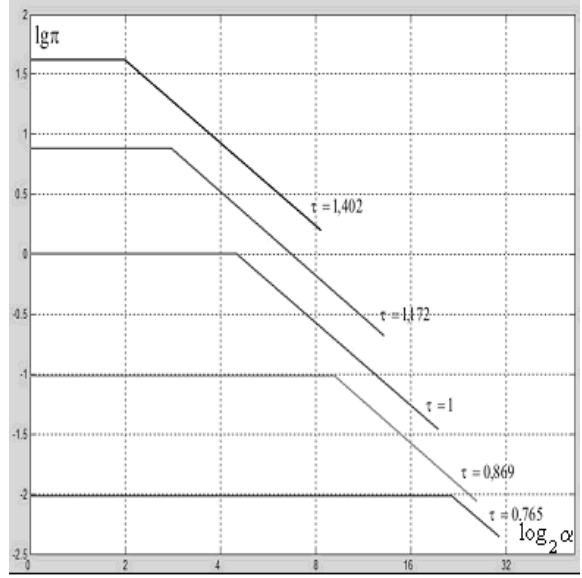

*Fig.1. Isotherms of real gas*

When matter is in a two-phase regime, then one of the terms comprising the grand partition function (6), has a characteristic maximum (Fig.2), i.e. one of the probability coefficients at a certain given value of $n = n_m$ reaches maximum value:

$$q_n^n = \left[\frac{N\alpha - n_m + 1}{n_m + 1} \cdot \tau^{i/2} \exp\left(r\left(1 - \tau^{-1}\right)\right)\right]^n \quad (9)$$

The corresponding term along with a small number of neighbouring polynomials makes a considerable contribution to the partition function (6). The function $f(n) = q_n^n$ declines monotonically on either side of the maximum.

In order to find the extremum of this function we need to define its logarithmic derivative. When we equate $d(\ln f(n))/dn$ to zero, we arrive at the transcendental equation:

$$\ln\left[\left(\frac{N\alpha}{n_m} - 1\right)A(\tau)\right] - \frac{N\alpha/n_m}{(N\alpha/n_m) - 1} = 0 \quad (10)$$

The root of this equation is a certain $D = N\alpha/n_m$, the magnitude of which depends on the temperature factor (5). If we fix the number of particles $N$ in the macrosystem, then parameter $n_m$ and specific volume $\alpha$ are proportional to each other such as:

$$n_m/\alpha = N/D. \quad (11)$$

There is a minimum number of particles in the gas phase

$$n_{\min} = N/D \quad (12)$$

that corresponds to value $\alpha = 1$, i.e. the critical volume.

When $n_m = N$, then from (11) we can determine the maximum value of specific volume at which the system enters the single-phase regime.

$$N/\alpha_{ãð} = N/D. \quad (13)$$

The table below show computed values for roots of equation (10) for certain given values of temperature factor $A(\tau)$ that are typical for thermodynamic parameters before, at and after the critical point.

Table

$n_m/N$ и $\alpha_{ãð}$ as a function of $A(\tau)$

| $A(\tau)$ | $(n_{m/N})_{\alpha=1}$ | $\alpha_{ãð}$ |
|---|---|---|
| $10^{-6}$ | 3,679 $10^{-7}$ | 2,718 $10^{6}$ |
| $10^{-5}$ | 3,679 $10^{-6}$ | 2,718 $10^{5}$ |
| $10^{-4}$ | 3,679 $10^{-5}$ | 2,718 $10^{4}$ |
| $10^{-3}$ | 3,676 $10^{-4}$ | 2,720 $10^{3}$ |
| $10^{-2}$ | 3,652 $10^{-3}$ | 2,738 $10^{2}$ |
| $10^{-1}$ | 3,429 $10^{-2}$ | 2,917 $10^{1}$ |
| 1 | 2,178 $10^{-1}$ | 4,591 |
| $10^{1}$ | 5,364 $10^{-1}$ | 1,864 |
| $10^{2}$ | 7,250 $10^{-1}$ | 1,379 |
| $10^{3}$ | 8,155 $10^{-1}$ | 1,226 |
| $10^{4}$ | 8,641 $10^{-1}$ | 1,157 |
| $10^{5}$ | 8,935 $10^{-1}$ | 1,119 |
| $10^{6}$ | 9,128 $10^{-1}$ | 1,096 |

The probability distribution function (8) exhibits interesting behaviour at the point of phase transition. Fig. 2 displays on the logarithmic scale dependence of $w(n)$ on the number of particles $n$ at different values of specific volume $\alpha$ from $\alpha = 1$ to $\alpha_{гр}$, as computed from (8). The total number of particles in both phases was taken as equal to $N = 10^{3}$. We have used the same values for specific temperature as when computing real gas isotherms for Fig.1. Parameter $r$ that is connected to the heat of vaporization for a given substance practically does not depend on temperature and is taken as constant. Most often its values for different substances span the range $3 < r < 7,5$. For our calculations we have taken $r = 5$.

It can be observed from Fig.2 that at the point of phase transition as the specific volume $\alpha$ grows the height of bell-shaped maxima of $w(n)$ decreases as $w(n) \sim \alpha^{-1/2}$. At the same time the half-width of the peaks grows as $\Delta n_{1/2} \sim \alpha^{1/2}$. These parameters depend on the number of $N$ at a given value of specific volume – $w(n_m) \approx N^{-1/2}$ and $\Delta n_{1/2} \approx N^{1/2}$. At the intersection of two- and single phase regions of the isotherms, i.e. the region that corresponds to the roots of equation (10), the function $w(n)$ looks like the left-hand side of the Gaussian distribution. At that point the maximum of probability is at $n = N$, the height of the maximum begins to grow and the width of the distribution function diminishes. The area under the curve remains constant and equal to one as it should according to the rule of normalized probability $\int_0^N w(n)\,dn = 1$. Although the value of $N$ is high under normal conditions, the upper limit of the integral should always be finite and correspond to the mass of the matter that consists of the two coexisting phases.

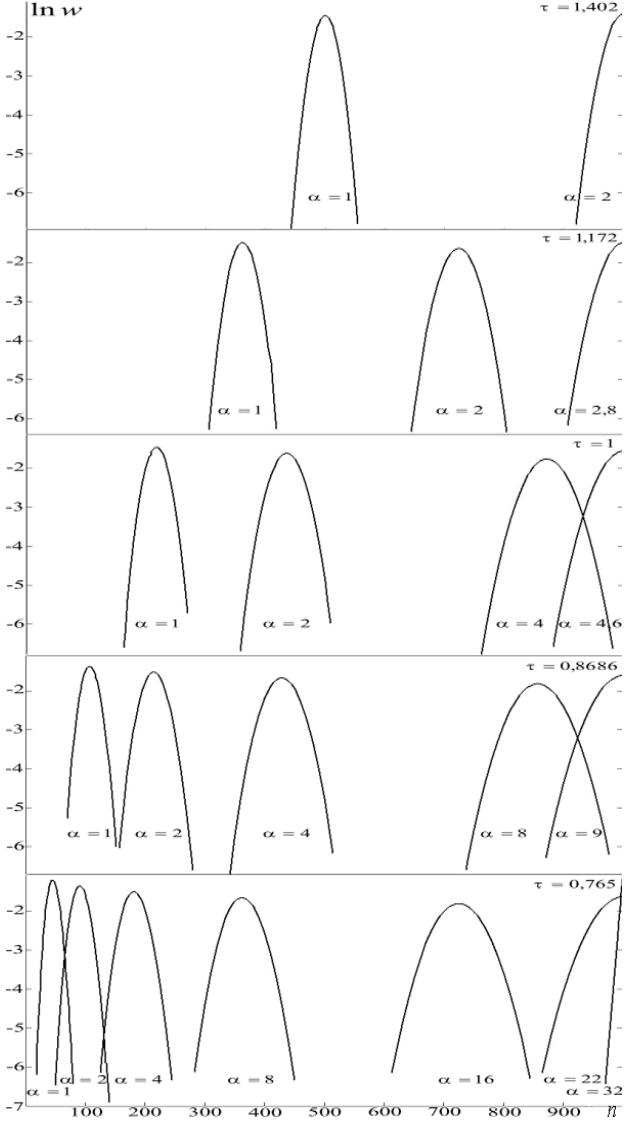

*Fig. 2. Probability of vaporization as a function of the number of particles for different values of specific volume and temperature.*

The computation of probability coefficients (9), partition function (6), mathematical probability (8), and hence specific equation of state (7) requires use of multiplicative terms that in turn create the need to handle computational problems arising from the use of large numbers.

The numerical problems can be solved if in the above expressions the coefficients $q_n^n$ are scaled by the largest of them

$$\widetilde{q}_n^n = q_n^n \big/ q_{n_m}^{n_m} \qquad (14)$$

making the modified probability coefficients $0 \le \widetilde{q}_n^n \le 1$. Similarly we can introduce modified partition function as

$$\widetilde{Z} = \sum_0^N \widetilde{q}_n^n.$$

It can be shown that already at $N \geq 10^4$ the modified probability coefficient (14) is in fact the exponent of the Gaussian distribution

$$\widetilde{q}_n^n \approx \exp\left(\frac{-(n-n_m)^2}{2\sigma^2}\right), \quad (15)$$

where $n_m$ – is the mean of the probability distribution; $\sigma$ is the distance from the mean to the inflection point of function (15); $\sigma^2$ is the variance.

In our case these values are

$$\sigma^2 = \langle n^2 \rangle - \langle n \rangle^2 = 0{,}1335\, N\alpha;$$
$$\sigma = 0{,}365 (N\alpha)^{1/2}. \quad (16)$$

When we introduce variable $x = (n - n_m)/\sigma$, the expression (15) can be reduced to the probability density $\varphi(x) = (1/\sqrt{2\pi})\exp(-x^2/2)$ of the normalized and centred probability distribution. Values of $\varphi(x)$ are tabulated in for example .

The modified partition function can now be computed by integration

$$\widetilde{Z} = \sum_0^N \widetilde{q}_n^n \cong \int_0^N \widetilde{q}_n^n\, dn = \int_0^N \exp\left(\frac{-(n-n_m)^2}{2\sigma^2}\right) dn.$$

For the last integral we can carry out a variable substitution analogous to the one that was done in expression (15). In that case the modified partition function can be expressed as the sum of two terms, each of which is an error integral

$$\widetilde{Z} = \sqrt{2\pi}\,\sigma\left[\hat{O}_0(ó_1) + \hat{O}_0(ó_2)\right], \quad (17)$$

where $\hat{O}_i(ó) = \int_0^ó \varphi(t)\, dt = (1/\sqrt{2\pi}) \times$

$$\times \int_0^y \exp(-t^2/2)\, dt = \sqrt{2\pi}\left[\hat{O}_0(ó_1) - \hat{O}_0(ó_2)\right];$$

$y_1 = n_m/\sigma; \quad y_2 = (N-n_m)/\sigma.$

The values of function $\hat{O}_0(ó)$ are also tabulated in .

Now the probability of vaporization (8) at $n_m < N$ becomes

$$w(n) = \frac{1}{\sigma} \cdot \frac{\varphi(x)}{\hat{O}_0(ó_1) + \hat{O}_0(ó_2)}$$

For the critical isotherm the maximal probability coefficient coincides with the last term of the polynomial

$$q_n^n \approx \left(\frac{N\alpha - N}{N+1}\right)^N = (\alpha - 1)^N.$$

The largest contribution to the partition function comes from the last terms of the polynomial.

In the middle of the two-phase region $\hat{O}_0(ó_1) + \hat{O}_0(ó_2) \cong 1$ and the modified partition function (17) is

$$\widetilde{Z} = \sqrt{2\pi}\,\sigma.$$

The mean of the Gaussian distribution for the case of $N = 10^{24}$ particles can be expressed using (17) in the following form:

$$w(n) = \frac{\widetilde{q}_{n_m}^{n_m}}{\widetilde{Z}} = \frac{1}{\sqrt{2\pi} \cdot \sqrt{N\alpha}} \sim 10^{-12}.$$

Using the above considerations we can write down the thermal equation of state of real gas (7) as follows

$$\pi = K\tau \left\langle \frac{n}{N\alpha - n + 1}\right\rangle \cong \frac{K\tau}{\sqrt{2\pi}} \frac{n_m/(N\alpha)}{1 - n_m/(N\alpha)}. \quad (19)$$

We can now conclude that use of modified probability coefficients and partition functions to calculate the pressure of real gas can circumvent numerical problems that arise in calculations of thermodynamic properties due to large num-bers. Hence the problem of deriving thermal equation of state for real gases can now be considered as solved in principle.

It is instructive to analyse the obtained result in the general framework of equations of state in thermodynamics.

The above examples illustrate that the thermodynamic methods can be successfully employed to tackle fundamental problems in natural sciences in general as well as in the applied fields such as geology and mining.

**Acknowledgement**

The authors are grateful to Dr Tetyana Bogdan and the University of Oxford for all the help in preparation of this manuscript.